\documentclass[11pt]{article}

\usepackage{fullpage,xr,graphicx,amssymb,amsfonts,amsmath}

\title{Identifying and Mapping Cell-type Specific \\ Chromatin Programming of Gene Expression}

\author{Troels T. Marstrand$^{1}$ and John D. Storey$^{1,2 *}$ \\
\\
\normalsize{$^{1}$Lewis-Sigler Institute for Integrative Genomics}\\
\normalsize{$^{2}$Department of Molecular Biology}\\
\normalsize{Princeton University, Princeton, NJ 08544, USA.}\\
\normalsize{$^{*}$Corresponding author: {\tt jstorey@princeton.edu}}
}

\date{}

\begin{document}

\maketitle

\begin{abstract}
A problem of substantial interest is to systematically map variation in chromatin structure to gene expression regulation across conditions, environments, or differentiated cell types. We developed and applied a quantitative framework for determining the existence, strength, and type of relationship between high-resolution chromatin structure in terms of DNaseI hypersensitivity (DHS) and genome-wide gene expression levels in 20 diverse human cell lines.  We show that $\sim$25\% of genes show cell-type specific expression explained by alterations in chromatin structure. We find that distal regions of chromatin structure (e.g., $\pm$200kb) capture more genes with this relationship than local regions (e.g., $\pm$2.5kb), yet the local regions show a more pronounced effect. By exploiting variation across cell-types, we were capable of pinpointing the most likely hypersensitive sites related to cell-type specific expression, which we show have a range of contextual usages. This quantitative framework is likely applicable to other settings aimed at relating continuous genomic measurements to gene expression variation.
\end{abstract}

\

\noindent Abbreviations: ARS, Angle Ratio Statistic; DHS, DNaseI hypersensitivity; SI, Supplementary Information

\ 

\noindent Note: The Supplementary Information may be found among the source files in arxiv\_SI.pdf.

\ 

\clearpage
\section{Introduction} Humans, like all other multicellular organisms, possess a large number of distinct cell-types, each of which is specialized for a particular function within the body.  Cells from a variety of tissue types exhibit different gene expression profiles relating to their function, where typically only a fraction of the genome is expressed. As all somatic cells share the same genome, specialization is in part achieved by physically sequestering regions containing non-essential genes into heterochromatin structures. Genes which are needed for the particular task of the cell-type display an accessible chromatin structure allowing for the binding of transcription factors and other related DNA machinery and subsequent gene expression. 

To date, most studies have been limited to considering the chromatin accessibility surrounding the promoter region of genes, which is typically proximal ($<$10kb) to the transcription region in just one or very few cell-types or experimental conditions \cite{Xi:2007:PLoS-Genet:17708682, Boyle:2008:Cell:18243105, Song:2005:Proc-Natl-Acad-Sci-U-S-A:15728362} .  However, it is also of interest to understand how larger regions ($\gg$10kb) of chromatin structure relate to a gene's expression variation across multiple cell types, disease states, or environmental conditions.  Recently, several large-scale international collaborations have started to generate data that can be used for this purpose \cite{Satterlee:2010:Nat-Biotechnol:20944594}, although doing so requires new developments in computational methods \cite{Hawkins:2010:Nat-Rev-Genet:20531367,Heintzman:2009:Nature:19295514,Ernst:2010:Nat-Biotechnol:20657582}.

To this end, we undertook a genome-wide investigation to characterize the relationship between variations in chromatin structure and gene expression levels across 20 diverse human cell lines (SI, Table~\ref{table:Celltypes}). We utilized data on chromatin structure as ascertained through DNaseI hypersensitivity (DHS) measured by next-generation deep sequencing technology, and gene expression data measured by Affymetrix exon arrays. Replicated data on 10 cell lines were also utilized to assess the robustness of our method.

\begin{figure}[p]
\centerline{\includegraphics[width=0.8\textwidth]{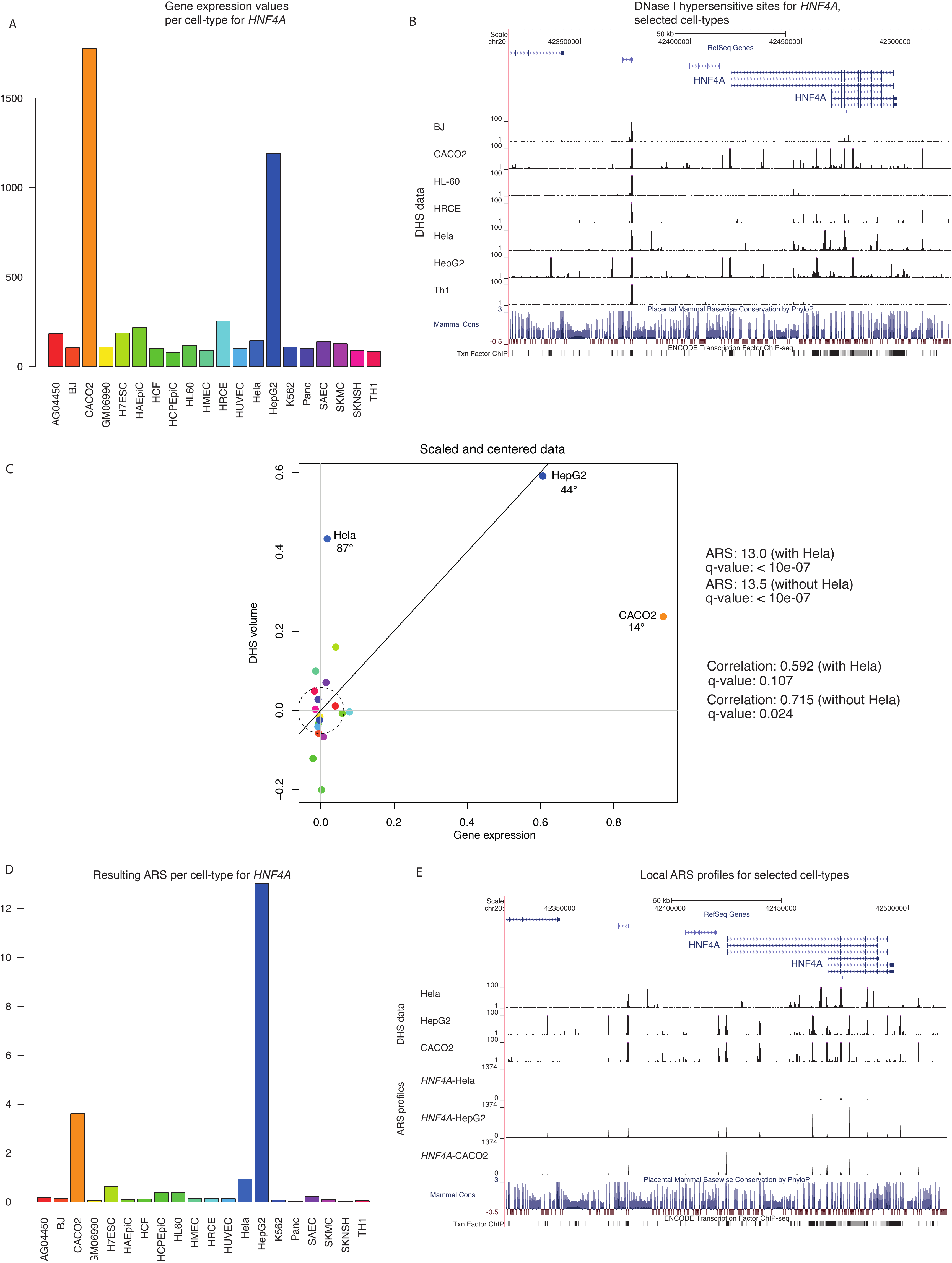}}
\caption{\scriptsize  \textbf{Overview of Data and Proposed Approach.} (A) Gene expression measurements for twenty cell lines on an example gene, {\em HNF4A}. (B) DNase-I Hypersensitivity (DHS) fragment sequencing counts in a region about the gene. (C) The DHS signal is captured by summing the overall number of fragments over a given segment size (e.g., $\pm$ 100kb) about the gene's transcriptional start sight (TSS) to obtain a ``DHS volume''.  After global normalization, the gene expression data and DHS volume measures are scaled to lie on the unit interval [0,1] and the data are centered about the origin according to the two-dimensional medoid.  For the {\em HNF4A} example, three outliers are clearly visible; for example, HepG2 displays both chromatin accessibility and active gene expression, whereas Hela displays only chromatin accessibilty.  The goal is to quantitatively capture the isolated relationship seen in HepG2 and assess whether this relationship is statistically significant.  Traditional measures of linear correlation are not suitable for identifying this type of signal, as shown by the substantial change seen after removal of a single cell line, Hela, even though the data for Hela are expected to exist for many genes and cell lines. The proposed ARS is robust to Hela since the measure is based on angular placement and the median distance to the medoid of the data (dashed circle). (D) The ARS statistic is calculating by first quantifying the relative distance to the origin for each cell line in a robust manner.  An angular penalty for each cell line is then calculated to quantify cell-types concordant in both expression and DHS measured. This quantity is measured in terms of angular distance from the $45^{\circ}$ degree line, and it is then multiplied times its respective relative distance to give and overall score for each cell line.  The maximum score is taken as the statistic for the given gene, allowing a comparison across all genes.   (E) A local version of the ARS statistic we introduce can pinpoint DHS ``peaks'' contributing the most to the detected association. See main text for details on the proposed methods.}
\label{fig:rawData}
\label{fig:method}
\end{figure}

Relating DHS to gene expression levels across multiple cell-types is challenging because the DHS represents a continuous variable along the genome not bound to any specific region, and the relationship between DHS and gene expression is largely uncharacterized.  In order to exploit variation across cell-types and test for cell-type specific relationships between DHS and gene expression, the measurement units must be placed on a common scale, the continuous DHS measure associated to each gene in a well-defined manner, and all measurements considered simultaneously. Moreover, the chromatin and gene expression relationship may only manifest in a single cell-type, making standard measures of correlation between the two uninformative because their relationship is not linear over a continuous range, as shown in Fig.~\ref{fig:rawData} (further details in SI and Figs.~\ref{fig:correlation}-\ref{fig:Spearman_ARS_200}).

The computational approach developed here provides a powerful, tractable, and intuitive way of representing these data and capturing biologically informative relationships. We were able to characterize the level to which variation of chromatin accessibility is associated with gene expression variation in a cell-type specific manner. Within genomic segments of significant chromatin gene expression concordance, our methodology is further capable of pinpointing the most likely local sites related to the detected association. We show that such sites are context specific and can be shared across genes within a single cell-type or across several cell-types. Our quantitative framework has some generality in that it may be readily applied to associate any quantitative measure along the genome to gene expression variation. 

\section{Results}
\subsection{Genome-wide profiling of chromatin accessibility and gene expression} We utilized data on genome-wide, high-resolution chromatin accessibility measurements for 20 distinct human primary and culture cell lines that were obtained by an established sequencing-based method \cite{Sabo:2004:Proc-Natl-Acad-Sci-U-S-A:15550541}.  In principle, accessible ``open'' chromatin is cleaved by the non-specific endonuclease DNaseI, and the cleaved fragments are sequenced to provide a high-resolution, genome-wide map of DNaseI hypersensitivity (DHS) for every cell-type (SI, Table~\ref{table:DNaseI}). The interpretation of these data is that increased fragment counts within a region are indicative of greater chromatin accessibility. To investigate the impact of regional chromatin accessibility on gene expression variation, we likewise utilized genome-wide gene expression measurements in each cell line from Affymetrix exon arrays (SI, Table~\ref{table:GEO}).  A total of 19,215 genes were analyzed after preprocessing (Methods).

With these quantifications, we sought to characterize the relationship between chromatin accessibility and gene expression in a cell-type specific manner, summarized in Fig.~\ref{fig:method}. To this end, the cell-type specific chromatin profiles were quantified by integrating the DHS fragment counts over increasingly larger genomic segments relative to the gene of interest (SI, Fig.~\ref{fig:genomicSegments}) to obtain a cell-type specific regional DHS volume.  We selected a range of segments that were likely to encompass all proximal (TSS $\pm$2.5kb) and most distal regulatory elements (TSS $\pm$50kb, $\pm$100kb, $\pm$150kb, $\pm$200kb, $\pm$100kb minus proximal 2.5kb, and $\pm$200kb minus proximal 2.5kb). Additionally, to account for copy number variation and chromosome arm related effects, the obtained DHS volumes were scaled on either side of the centromere to arrive at equilibrium across samples (SI, Fig.~\ref{fig:DHSScaling}). Alternative representations of DHS signal \cite{Ernst:2010:Nat-Biotechnol:20657582,Day:2007:Bioinformatics:17384021} could be utilized at this step, although we did not identify any advantages in doing so. Gene expression values were summarized as the mean intensity across all probe-sets linked to a given RefSeq-gene.
 
\subsection{Detecting cell-type specific chromatin accessibility and gene expression concordance} 
\begin{figure}[p]
\centerline{\includegraphics[width=0.5\textheight]{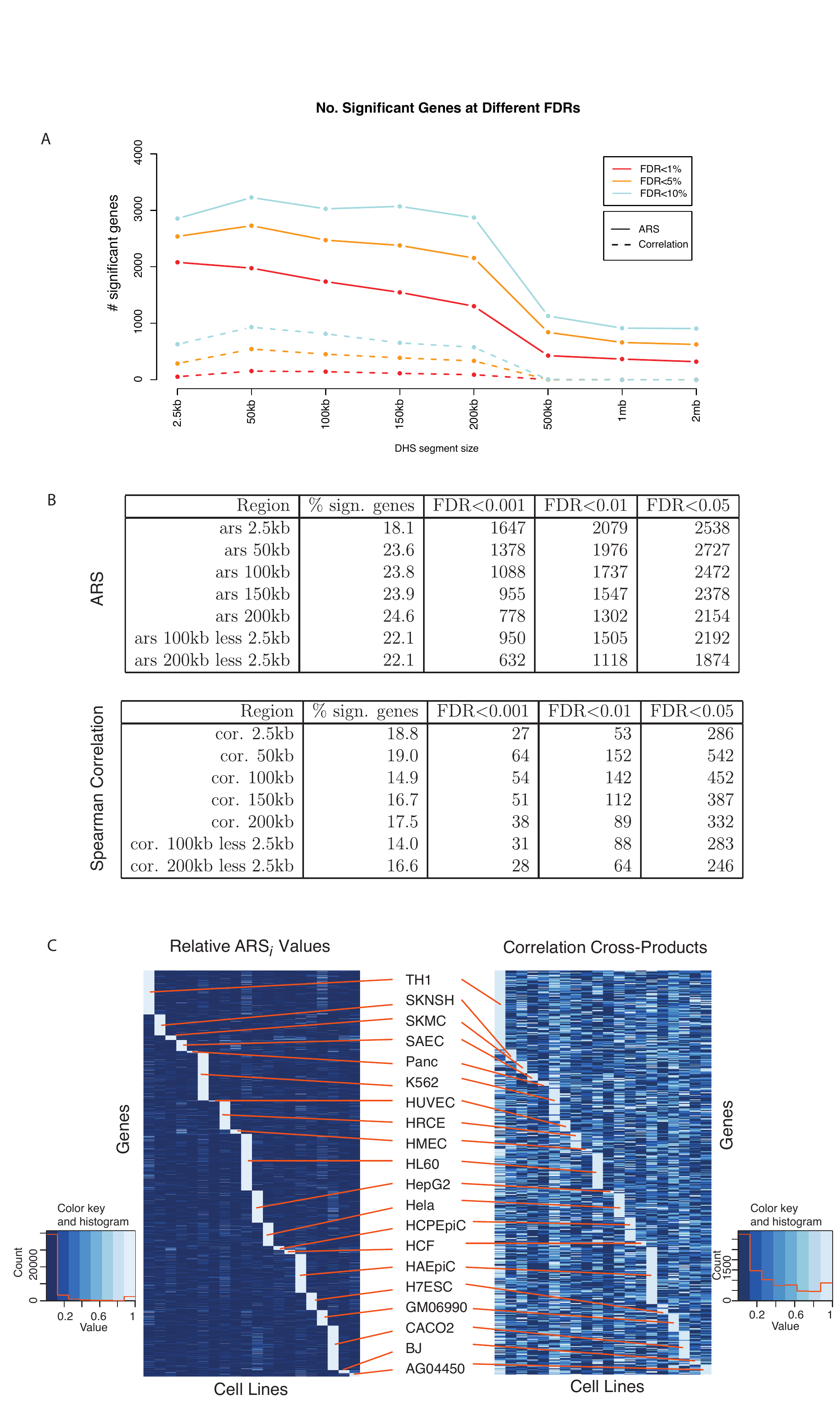}}
\caption{\scriptsize  \textbf{Statistical significance for ARS and correlation across genomic segments} (A) Depicts the number of significant genes found at increasingly larger genomic segments for ARS and Spearman correlation, respectively (solid line is ARS and dashed line is Spearman correlation). (B)  Statistical significance according to DHS volume segment size. Column 2 shows the percentage of genes estimated to have  concordant DHS volume and gene expression variation as captured by ${\rm ARS}_{\rm max}$ ($1-\hat{\pi}_0$, as estimated in \cite{Storey:2003:Proc-Natl-Acad-Sci-U-S-A:12883005}). Columns 3-5 show the number of statistically significant genes at various FDR cut-offs.  While the 2.5Kb window shows more significant genes at the stringent FDR cut-offs, indicating a larger effect size, the overall percentage of genes showing a relationship is notably lower than the more distal DHS volumes. Compared to Spearman correlation, ARS is more powerful at detecting these associations (see SI for further details). (C) The relative ARS$_i$ values across all cell-types for significant genes in the $\pm$100kb region versus the analogous components for Spearman correlation (the cross product terms that sum to form the overall correlation). The ARS$_i$ values distinguish cell lines that have a strong DHS and expression concordance substantially more clearly than the Spearman correlation, showing that the traditional correlation is more likely to generate spurious results from small changes to the data. Enrichment of biological functions for the significant genes found by either method corroborates this finding (see SI, Fig. \ref{fig:Ingenuity})}
\label{table:FDR}
\label{fig:relativeARS}
\end{figure}

Due to the ``on-off'' nature of DHS and subsequent transcription, there will not necessarily be a linear relationship between DHS and gene expression measures. Using correlation or correlation-like statistics to associate the two measurements across all cell-types proved to be unreliable and uninformative (further details in SI and Figs.~\ref{fig:correlation}-\ref{fig:Spearman_ARS_200}). One of the key types of relationships we sought to detect is of the type shown in Fig.~\ref{fig:rawData}, where one or very few cell types are outliers from the others. The standard Pearson correlation statistic is not well-suited for this scenario.  First, it requires the data to be jointly Normal in order to obtain parametric p-values, but the Normal assumption does not hold for these data (SI, Fig.~\ref{fig:Normal_test}).  Second, this correlation statistic is unstable when there are outliers, even when using permutation based p-values, demonstrated directly on these data (SI, Figs.~\ref{fig:correlation} and \ref{fig:Pearson_perm_pvalue}).  The rank-based Spearman correlation statistic is a potential alternative, but it shows very poor power relative to the method proposed here as shown in Fig.~\ref{fig:relativeARS}. (See also SI, Figs.~\ref{fig:Spearman_ARS_25} and \ref{fig:Spearman_ARS_200}).  For example, at a false discovery rate (FDR) $\leq 0.05$, the proposed method identifies 2538 genes with a cell-type specific DHS and gene expression relationship whereas the Spearman statistic identifies only 286 (Figs.~\ref{fig:relativeARS}, \ref{fig:Spearman_ARS_25}, and \ref{fig:Spearman_ARS_200}).  

The new statistic proposed here is designed to be appropriate for scenarios when both measurements are restricted to a narrow relative range with one or very few cell-types appearing as distinct outliers. To evaluate the relationship between the DHS volume of a genomic segment and gene expression, we took into account the compactness of the measurements versus any distinct outliers in both dimensions and whether the outliers were concordant in both measurements (i.e., a simultaneous increase or decrease)  to form an overall composite measure called an Angle Ratio Statistic (ARS) (detailed in Fig.~\ref{fig:method}, Fig.~\ref{fig:method_supplementary}, Methods, and SI).  To summarize, we first scale and median center the DHS volume and expression data, respectively, for a given gene.  We then calculate the relative distance of each cell type to the overall center of the data, which serves as a way to measure the degree to which each cell type is an outlier.  In order to measure concordance of DHS volume and gene expression, we calculate the angular distance between each point and the 45$^\circ$ line of identity, penalizing points further away from the line of identity according to a data-derived exponential function.  These two quantities are then multiplied to form an ARS$_i$ value for each cell type ($i=1,2,\ldots,20$), and the maximal value ARS$_{\max}$ is the overall statistic that quantifies cell-type specific DHS volume and gene expression concordance for a given gene.

To identify statistically significant genes from $\mathrm{ARS}_{\max}$, we constructed a null distribution based on randomization of the observed experimental data (see Methods, SI, and Fig.~\ref{fig:completeContour}). $\mathrm{ARS}_{\max}$ values obtained from the randomized data were used as a basis for determining a p-value of the observed $\mathrm{ARS}_{\max}$ for each gene. False-discovery rate (FDR) based statistical significance and the proportion of genes with a true chromatin accessibility and gene expression relationship were estimated from the p-values \cite{Storey:2003:Proc-Natl-Acad-Sci-U-S-A:12883005} (Figure~\ref{table:FDR}b). 
 
We estimate that $\sim$25\% of genes show concordance between chromatin accessibility and gene expression variation in a cell-type specific manner. While our strategy is capable of detecting outliers showing negative concordance (decreased chromatin accessibility and decreased gene expression), none were found to be significant at FDR $\leq$ 0.05. The number of significant genes increased by inclusion of distal DHS volume (Fig.~\ref{table:FDR}b, column 2), indicating that distal chromatin programming effects are more widespread in a genome-wide sense. On the other hand, using the proximal DHS volume we observe a greater empirical effect size compared to the distal DHS volumes (Fig.~\ref{table:FDR}b, columns 3-5). 

This observation is explained by the aggregation of genes significant for the same cell-type along the genome \cite{Sproul:2005:Nat-Rev-Genet:16160692}. Testing whether one or more significant genes within a $\pm$100kb region were associated with the same cell-type we found that 481 out of 668 significant genes within the specified boundary stem from the same cell-type (Fishers exact test p-value $<$ 2.2e-16; SI and Fig.~\ref{fig:aggregationQQ}). It is however important to note that inclusion of increasingly distal regions also increases the noise in the DHS volume, wherefore the effect size and ultimately the number of true associations starts to decline (Fig.~\ref{table:FDR}a).

\subsection{Experimental replication} To assess reproducibility, we tested the concordance of significant results among replicated data for 10 cell-types. Based on two independent measurements of DHS and gene expression, respectively, we calculated the fraction of predictions preserved in all four-way comparisons (SI). We found that between 86\% to 91\% of significant genes (FDR $\leq$ 0.05) were identical (SI, Fig.~\ref{fig:reproduce}).
  
\subsection{Gene ontology and pathway analysis} To determine the biological coherence of the set of genes found to be significant for each cell-type, we performed a gene ontology (GO) enrichment analysis \cite{Eden:2009:BMC-Bioinformatics:19192299}. The method computes enrichment within the process and function components of GO categories and assigns a numerical significance to the findings. In nearly all cases the results were in agreement with the actual biology; see {\tt http://encode.princeton.edu/} for results on all DHS segment sizes. For example, human T-cells showed a strong enrichment of T-cell receptor related genes, whereas hepatic cells showed enrichment of lipid metabolism related genes. KEGG pathways \cite{Kanehisa:2002:Novartis-Found-Symp:12539951, Dennis:2003:Genome-Biol:12734009} were likewise enriched in a cell-type specific manner. For example, HepG2 showed significant enrichment for genes within the bile acid synthesis and drug metabolism, while HL60 showed significant enrichment within the hematopoietic cell lineage (data not shown). 

Furthermore, all genes detected within each cell type at FDR $<$ 0.05 ($\pm$ 100kb DHS volume) were analyzed through the use of Ingenuity Pathways Analysis (Ingenuity Systems\textregistered, {\tt www.ingenuity.com}). For all but three cases out of 20 (two cell-types likely had too few significant genes detected to get reliable annotations), the category ``Physiological System Development and Function'' was in clear correspondence with that expected given the cell type, (SI, Fig.~\ref{fig:Ingenuity}). For instance, TH1 was enriched for ``cell-mediated immune response'', K562 for ``hematological system development and function'', and H7ESC for ``embryonic development''. For each gene, there tended to be low relative $\textrm{ARS}_i$ across the remaining cell types, indicating that we detected truly cell-type specific genes as clear outliers on a genome-wide scale. However, some cases showed large relative $\textrm{ARS}_i$ in a few tissues, which prompted us to investigate these instances further.

Among genes with a statistically significant ARS$_{\rm max}$ statistic, additional inspection of the remaining ARS$_i$ were explored for detection of possible sub-structures.  We calculated relative ARS values within each gene dividing all $\mathrm{ARS}_i$ by $\mathrm{ARS}_{\max}$. In addition to many instances of singular outliers, we detected a gradient behavior among significant genes, where a few cell-types were evident as outliers (Fig.~\ref{fig:relativeARSsupp}). 

\subsection{Local ARS Profiles}  The DHS data itself provides a rich source of information about regulatory elements in the genome.  However, when used in conjunction with gene expression data across differing cell types, there is an opportunity to discover which locations of chromatin accessibility drive gene expression in a cell-type specific manner.  This goal prompted us to develop a technique to model the relationship for fine-scale segments of DHS volume across the larger segments. As the above strategy focused on examination of chromatin gene expression interactions over genomic segments, investigation of fine-scale patterns allowed us to: (i) validate that distal regulatory regions were indeed present as peaks in chromatin accessibility  concordant with gene expression in a cell-type specific manner, (ii) perform sequence analyses of these chromatin accessibility peaks, (iii) compare localized associations across cell-types or within a single gene, and (iv) provide a framework for quantifying regions of interest on a continuous scale for investigation of regulatory elements.

We therefore extended our approach to allow one to identify and map DHS sites to genes on which they show strong evidence for playing a regulatory role in a cell-type specific manner.  This was carried out by providing a fine-scale version of the ARS quantification, called a ``local ARS profile'' for genes with a statistically significant $\mathrm{ARS}_{\max}$ statistic over a larger segment. The peaks of the local ARS profiles pinpoint which DHS are most influential in explaining the cell-type specific gene expression variation, thereby indicating that they have the most regulatory potential.  We retained the gene expression values for a given significant gene, and now considered the DHS volume within non-overlapping consecutive regions at a high resolution 60 base pair windows. The ARS statistic was calculated for each 60bp window, which can then be plotted over the entire region used in identifying the gene as statistical significant.  For example, for a gene significant with respect to a $\pm$200kb DHS volume, we calculated $\sim$6700 local ARS statistics for each cell type.  These can then be plotted in such a way that the signal emanating from that location is visible, loosely analogous to a LOD score profile in linkage analysis.  Additional steps were taken, involving scaling across the 60 bp windows to preserve a valid interpretation of their relative magnitudes (SI).  

We first selected the subset of local ARS profile ``peaks'' by thresholding the local ARS profiles in a principled manner (Methods), and we analyzed both positional biases and sequence compositions as they relate to function.  We then analyzed the entire trajectories of local ARS profiles at specific loci, showing that they identify both known and putative regulatory DHS for given genes. 

\begin{figure}[p]
\centerline{\includegraphics[width=0.8\textwidth]{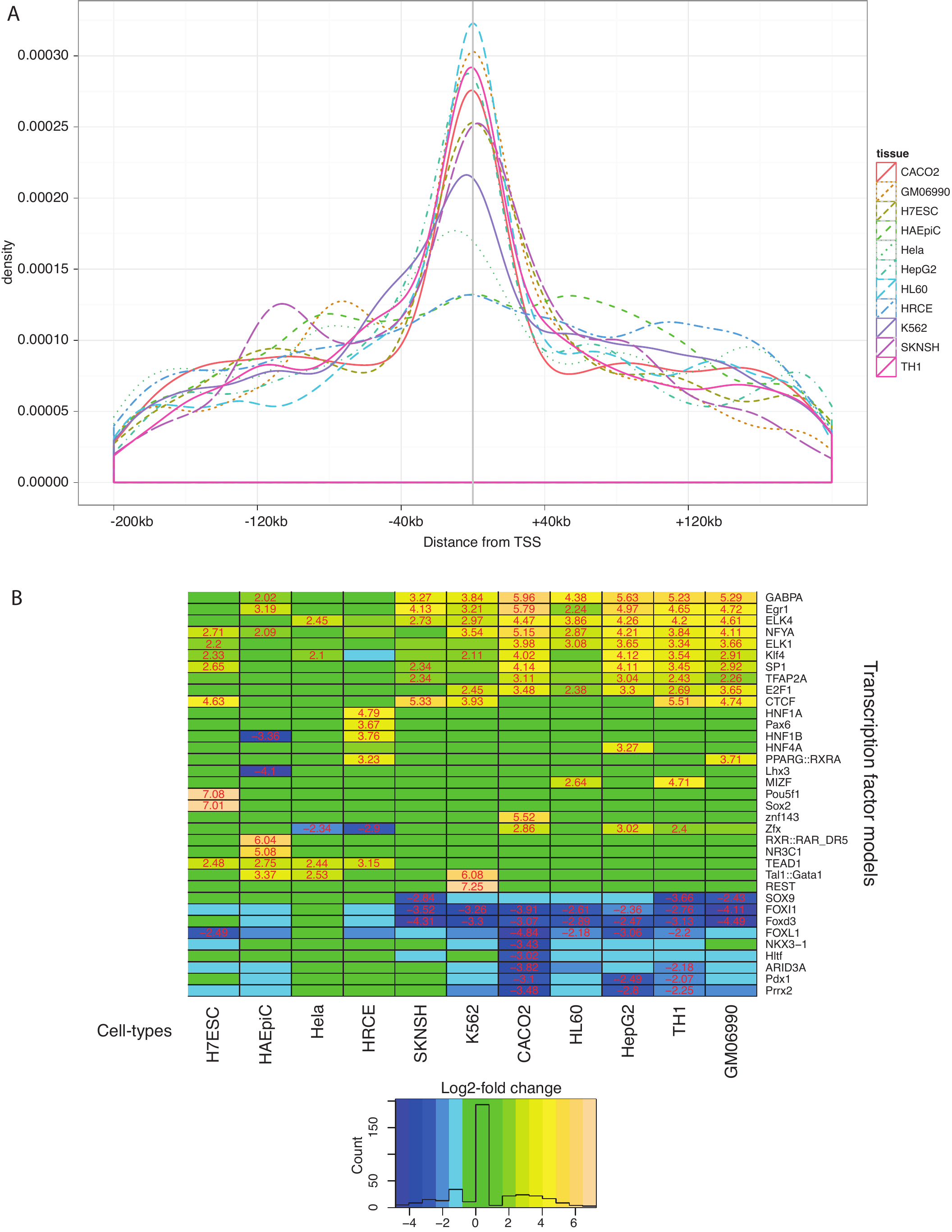}}
\caption{\footnotesize  {\bf Analysis of local ARS profiles.} (A) {\em Distribution of local ARS peaks relative to the TSS according to cell type.} The positional bias of cell-type specific local ARS peaks as measured by the density of local ARS peaks within cell lines with respect to position from to the TSS. Clear differences in the amount of distal regulation are seen across the cell-types and the density around the TSS differ markedly among cell types. For example, HL60 shows a more proximal signal relative to that of HAEpiC.  (B) {\em Transcription factor binding site analysis among local ARS peaks occurring 10kb to 200kb from the TSS.} Sequences corresponding to local ARS peaks within significant cell-type specific genes were searched with known transcription factor binding site models, and the relative over- and under-representation was assessed based on a negative control set. Instances of absolute $\log_2$ fold-change $\geq 2$ are displayed within the relevant cell types. Over-representation is indicative of a preferential transcription factor binding site, and is therefore a likely regulatory candidate for the observed gene expression. Under-represented sites indicate factors which should be avoided to maintain proper cell-type specific expression profiles. For instance, Sox2 and Pou5f1 (Oct4) were observed solely over-represented in the embryonic cells, H7ESC.}
\label{fig:newbio}
\end{figure}

\subsection{Positional bias of putative regulatory DHS}
Because the overall statistical significance increases when calculating DHS volume over more distal regions up to 200kb (Fig.~\ref{table:FDR}), we investigated the positional bias of local ARS peaks in a cell-type specific manner. Figure \ref{fig:newbio}a shows smoothed densities of positional local ARS peak counts by cell type, which exhibit high cell-type specific differences, specifically the density around the TSS.  Random densities were generated by randomly assigning positional counts to tissues in equal proportions to the observed counts, where it can be seen that the cell-type differences are no longer present (SI, Fig.~\ref{fig:distalARS}). This points to the existence of cell-type specific biases in the base-pair distance of regulatory DHS to TSS.

\subsection{Sequence analysis of peaks in local ARS profiles}
We next sought to characterize the functional significance of sequences corresponding to local ARS peaks.  Since a general indicator of functionality is conservation, we extracted the conservation track values (phastCons44wayPrimate, hg18) \cite{Siepel:2005:Genome-Res:16024819} corresponding to the local ARS peaks and to the negative control set (Methods). Values range from 0 to 1, with 1 indicating the most conserved. The regions with local ARS peaks were significantly more conserved than regions from the negative control set (Kolmogorov-Smirnov p-value $<$ 2.2e-16, SI, Fig.~\ref{fig:qq_cons}), indicating substantial conservation of sequences corresponding to local ARS peaks.  \label{sup-para:distal}

DNase-I hypersensitive sites are well established markers of regulatory and other DNA binding proteins. We therefore sought to establish if known cell-type specific transcription factors binding sites (TFBSs) are over-represented in the local ARS peaks relative to the negative control set (Methods).  Since regions distal to the TSS are rarely studied in this context, we eliminated all local ARS peaks and negative controls that fell within $\pm$10 kb of the TSS.  This step was taken to demonstrate that the proposed approach is capable of detecting distal TFBS, up to 200kb from the TSS.  

We utilized the JASPAR database \cite{PortalesCasamar:2010:Nucleic-Acids-Res:19906716} to identify TFBS that are differentially represented in the local ARS peaks relative to the negative control set (Methods).  The over- and under-represented TFBS show distinct cell-type specific patterns and provide a rich insight into cell-type specific gene regulation (Fig.~\ref{fig:newbio}b), several of which are listed here:
\begin{itemize}
\item Among the hepatocyte nuclear factors we found {\em HNF1B} (TCF-2) and {\em HNF4A} to have significant chromatin gene expression concordance in HRCE and HepG2, respectively (SI, Fig.~\ref{fig:HNF1B} and Fig.~\ref{fig:HNF4A}). Furthermore we found the local ARS profiles in the respective tissues to display a marked over-representation of the factor in question, {\em HNF1B} in HRCE and {\em HNF4A} in HepG2. Mutations in {\em HNF1B} have been associated with a broad range of renal diseases \cite{Ulinski:2006:J-Am-Soc-Nephrol:16371430}, and {\em HNF4A} is essential for hepatocyte differentiation and morphology \cite{Parviz:2003:Nat-Genet:12808453}.
\item H7ESC was found to show over-representation of {\em SOX2} and {\em POU5F1} (Oct-4) both essential for self-renewal in undifferentiated stem cells.
\item {\em NFYA} (a CCAAT-binding protein) was found over-represented in almost all tissues. This factor is essential for enhancer function by requiting distal transcription factors to the proximal promoter region  \cite{Wright:1994:EMBO-J:8076600}. The ubiquitous CCAAT-binding factor family is linked to cellular differentiation in a variety of tissues \cite{LekstromHimes:1998:J-Biol-Chem:9786841}.
\item RXR-RAR was found in HAEpiC (human amniotic epithelial cells). The co-expression of the retinoic acid receptors (RARs) and the retinoid X receptors (RXRs) \cite{Sapin:1997:Dev-Dyn:9022057} are essential for proper placental development, and retinoid X receptor (RXR) null mouse mutants are lethal after 10 days due to placental defects \cite{Sapin:1997:Dev-Biol:9356169}. 
\item Forkhead binding sites were found to be primarily under-represented, specifically {\em FOXD3} was under-represented in, among others, the leukemic cell-types. Silencing of {\em FOXD3} by aberrant chromatin modification has been implicated in leukemogenesis \cite{Sapin:1997:Dev-Dyn:9022057}. Over-expression of {\em FOXD3} prevents neural crest formation \cite{Pohl:2001:Mech-Dev:11335115}. Interestingly, binding sites for the factor were under-represented in SKNSH, a neuroblastoma derived from neural crest cells.
\item NF-$\kappa$B was found over-represented in TH1, where it promotes the expression of, among others, interleukin 12 (IL-12) essential for TH1 development \cite{Murphy:1995:Mol-Cell-Biol:7565674}.
\end{itemize}
The differentially represented TFBSs were distributed largely distal. For all cell-types, from 68\% to 79\% were located more than $\pm$50kb away from the TSS. We repeated the analysis with only the proximal regions ($\pm$10kb from the TSS), and we found that important known cell-type specific motifs were no longer detected (SI, Fig.~\ref{fig:prox_heatmap}).

\begin{figure}[p]
\centerline{\includegraphics[width=0.8\textwidth]{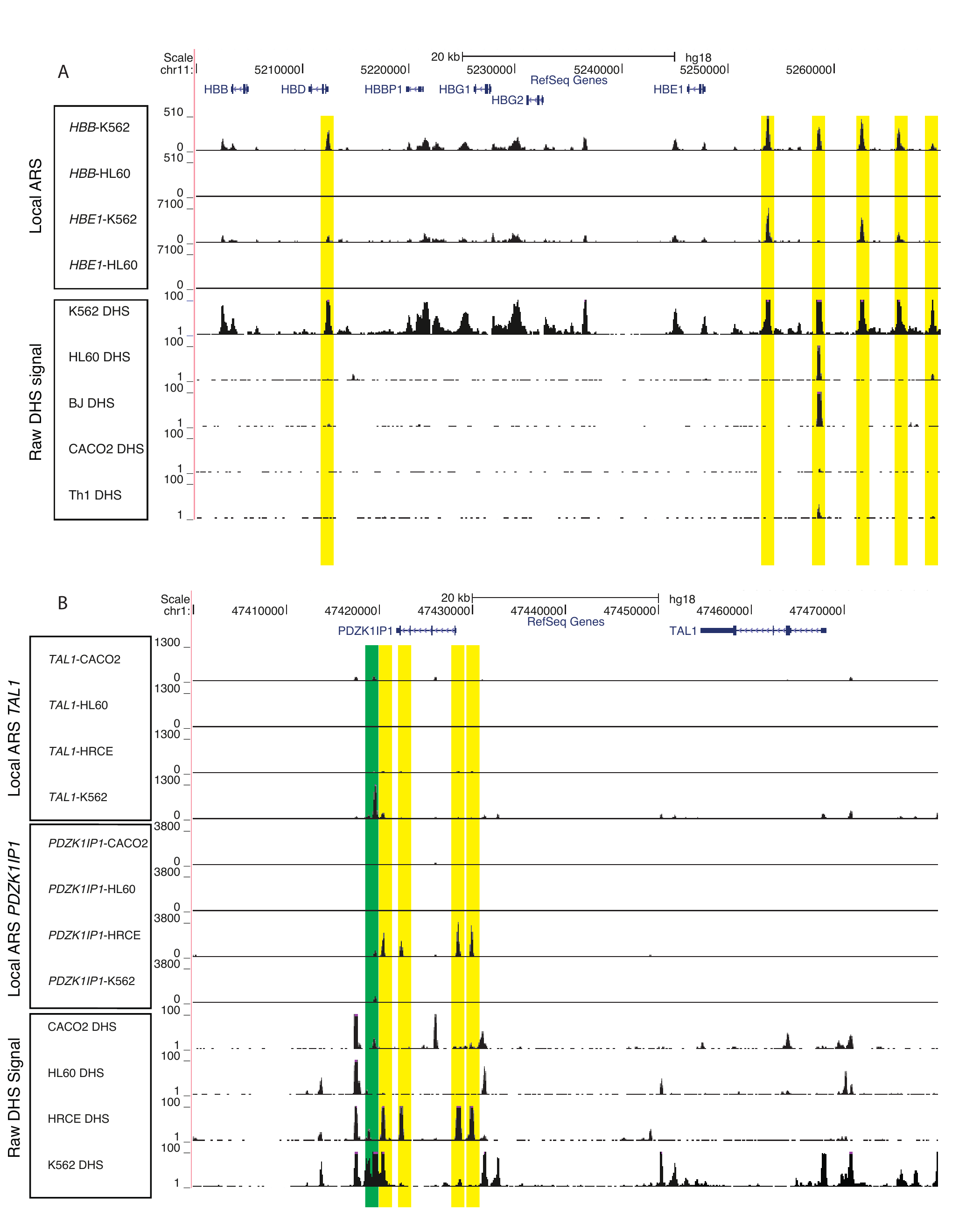}}
\caption{\footnotesize {\bf Mapping putative regulatory DHS with local ARS profiles at two loci.} (A) {\em $\beta$-globin locus control region.} The DHS data for five cell lines (out of 20) are shown, as well as the local ARS profiles for \emph{HBB} and \emph{HBE1} in the K562 and HL60 cell lines. The transparent yellow boxes indicate regulatory regions, specifically hypersensitive regions 1-5 (HS1-5), together with a less characterized site upstream of \emph{HBD}. It can be seen that \emph{HBB} and \emph{HBE1} show different local ARS profiles, indicative of differences in usage of regulatory elements. The local ARS profile shows no peak in HL60 despite the existence of a hyper-sensitive site when considering DNaseI profile alone.  The full data and local ARS profiles for all 20 cell lines and both genes are displayed in Figs.~\ref{fig:HBB_DHS}-\ref{fig:HBE1_ARS}.
(B) {\em TAL1 locus.} 
We identified \emph{TAL1} as statistically significant with its maximal ARS in the K562 cell line across all tested genomic segments. Local ARS profiles show a dominant effect from the +40 enhancer region (green box), spanning \emph{PDZK1IP1}. DHS signals across multiple cell-types were correctly not detected to be associated with the expression of \emph{TAL1}. Furthermore, note that even though the DHS data for {\em TAL1} and {\em PDZK1IP1} are largely overlapping, they nevertheless have distinct local ARS profiles due to their different patterns of gene expression.  This demonstrates that ARS is capable of separating interwoven signals across cell-types for neighboring genes, and that there is information to be gained by combining DHS and gene expression profiling. The full data for all 20 cell lines and local ARS profiles are displayed in Figs.~\ref{fig:TAL1_DHS}-\ref{fig:TAL1_ARS}.}
\label{fig:HBB}
\label{fig:TAL1}
\end{figure}

\subsection{Mapping putative regulatory DHS to genes}  
We also investigated the utility of considering the entire trajectory of local ARS profiles at a locus to characterize the regulatory architecture of cell-type specific expression. We investigated in detail two well-characterized examples of regulatory interactions at the $\beta$-globin (\emph{HBB}) locus control region and at the stem cell leukemia (\emph{SCL}) gene, also known as \emph{TAL1}, with several more appearing in SI (Figs.~\ref{fig:HBB_DHS}-\ref{fig:LOXL2}).  It can be seen from these analyses that the local ARS profiles provide a means to map DHS sites to genes in a cell-type specific manner.

The \emph{HBB} ($\beta$-globin) locus control region (LCR) comprises an array of functional elements that {\em in vivo} gives rise to five major DNase I hypersensitive sites (HS1-HS5 \cite{Tuan:1985:Proc-Natl-Acad-Sci-U-S-A:3879975, Forrester:1986:Proc-Natl-Acad-Sci-U-S-A:3456593, Grosveld:1990:Ann-N-Y-Acad-Sci:2291544}, Fig.~\ref{fig:HBB}) upstream of \emph{HBE1} ($\epsilon$-globin) on the short arm of chromosome 11.  All five sites were present in cell line K562 according to our DHS data (see Figs.~\ref{fig:HBB_DHS}-~\ref{fig:HBE1_ARS} for complete data across all 20 cell-types, and Fig.~\ref{fig:HBB_all} for local ARS profiles across all genes at this locus control region).  Although the DHS volume at these sites contributed to both \emph{HBE1} and \emph{HBB} yielding statistically significant ARS values, the relative importance of HS1-5 differs significantly between these two genes, clearly detected by the local ARS profiles (Fig.~\ref{fig:HBB}a).  

In the case of \emph{HBE1}, we observed local ARS peaks for HS1 at -6.1kb and to a lesser extent HS3 and HS4 (-14.7 and -18kb respectively). For \emph{HBB} we observed similar local ARS profiles for HS1, HS3 and HS4, and smaller local ARS values for HS2 (-10.9kb). It has previously been shown that HS1 is a stable chromatin structure \cite{Forrester:1986:Proc-Natl-Acad-Sci-U-S-A:3456593} through out development and essential for \emph{HBE1} expression \cite{Shimotsuma:2010:J-Biol-Chem:20231293} due to a GATA-1 binding site, while HS2, 3 and 4 show synergistic enhancement of expression of \emph{HBB} by formation of the LCR holocomplex \cite{Fraser:1990:Nucleic-Acids-Res:2362805, Molete:2001:Mol-Cell-Biol:11287603, Tolhuis:2002:Mol-Cell:12504019}. Finally, the element upstream of \emph{HBD} has also been reported to specifically enhance transcription of \emph{HBB} \cite{Acuto:1996:Gene:8654951}. While HS5 is present in the DHS data for K562, similar open chromatin structures were detected in other tissues. HS5 (-21kb) is not in concordance with tissue-specific gene expression of either \emph{HBE1} or \emph{HBB}, an observation in line with this site's function as an insulator and CTCF binding site \cite{Tanimoto:2003:Mol-Cell-Biol:14645507}.

\emph{TAL1} encodes a basic helix-loop-helix protein which is essential for the formation of all hematopoietic lineages (SI, Figs. ~\ref{fig:TAL1_DHS}-~\ref{fig:TAL1_ARS} for all data across the 20 cell-types). Previous studies using chromatin structure, comparative sequence analysis, transfection assays \cite{Fordham:1999:Leukemia:10374880, Gottgens:2002:EMBO-J:12065417}, and transgenic mice  \cite{Chapman:2003:Genomics:12659809, Gottgens:2001:Genome-Res:11156618, Gottgens:2002:Genome-Res:11997341, Sanchez:1999:Development:10433917, Sinclair:1999:Dev-Biol:10208748} have identified a total of five enhancers modulating the expression of \emph{TAL1}. We detect \emph{TAL1} as significant with maximal cell line K562 across all tested genomic segments (from $\pm$2.5KB to $\pm$200KB) with the most significant $\mathrm{ARS}_{\max}$ occurring for $\pm$50kb. Further investigation by the local ARS profile (Fig. \ref{fig:TAL1}b) showed that while proximal regulatory sites were correctly identified, the most dominant signal is by far confined to the +40 enhancer region and is an order of magnitude greater than other signals. While the \emph{TAL1} +40 region is downstream of \emph{PDZK1IP1}, it was not linked to the expression of this gene which was detected as significant in HRCE. The +40 enhancer region has been shown to direct expression in transgenic mice to primitive, but not definitive erythoblasts, such as the phenotype displayed by K562. This example demonstrates that our methodology is capable of identifying regions of regulatory potential, which otherwise requires laborious effort to annotate.

Local ARS profiles showed both differences and similarities across genes as well as cell-types. A few examples included: 
\begin{itemize}
\item \emph{CCR2} and \emph{CCR5} were significant for two different cell-types, HL60 and TH1, respectively (SI, Fig.~\ref{fig:CCR}).
\item Part of the HOX-cluster crucial for kidney development in mammals (\emph{HOXD8}, \emph{HOXD4}, and \emph{HOXD3}) showed identical local ARS profiles (SI, Fig.~\ref{fig:HOXD}),  and all were significant genes in HRCE \cite{DiPoi:2007:PLoS-Genet:18159948}. 
\item Another example of shared profiles, but across several cell-types instead of across several genes, was seen with \emph{LOXL2}, a gene essential for biogenesis of connective tissue, which is detected as an outlier in SKMC and has high relative ARS values in HAEpiC and BJ (SI, Fig.~\ref{fig:2DLOXL2}). Further fine-scale investigation showed a solid overlap in the local ARS profiles (Fig.~\ref{fig:LOXL2}). 
\end{itemize}
These observations point to a potentially widespread sharing of regulatory mechanisms both across genes and cell-types. 

\section{Discussion} As the epigenome in multicellular organisms is a dynamic entity whose variation leads to reprogramming of gene expression \cite{Wong:2005:Hum-Mol-Genet:15809262}, it is a likely candidate in the etiology of disease complementary to that of mutations in DNA \cite{Schneider:2010:Nucleic-Acids-Res:20194112, Hatchwell:2007:Trends-Genet:17953999}. It is therefore of considerable interest to identify and characterize the regulatory regions contributing to gene expression variation with respect to a given disease.

We have presented a framework for quantifying relationships between chromatin structure and gene expression across multiple conditions (here, cell-types),  facilitating a new avenue for understanding cellular responses by localizing and characterizing regions of regulatory potential. The local ARS profiles we introduced allow specific hypersensitive regions to be associated with condition-specific gene expression, thereby conferring contextual regulatory information not obtainable using DHS data alone. This effectively pinpoints a shortlist of primary candidates for further functional studies.  We found the peaks from the local ARS profiles in statistically significant segments to be both highly conserved and enriched for known transcription factor binding sites as far as 200kb from transcription start sites. While beyond the scope of the current work, we believe our approach could be used in conjunction with quantitative trait analyses (QTL) to increase the power for detecting true cis- and trans-acting SNP by interfering with transcription factor binding sites which in turn leads to altered DHS signals in a similar manner as Degner et al. \cite{Degner:2012fo}.

As measurements from high throughput sequencing platforms become commonplace in molecular biology, there will be an increasing demand for the development of new statistical approaches for these data.  A major challenge is that sequencing measurements are rarely in units directly relatable to one another; e.g., DHS measures chromatin accessibility, ChIP-seq measures binding affinity, RNA-seq measures RNA molecule abundance, etc. Our framework provides the initial development of a statistic which captures relationships among these measurements and enables statistical testing of associations among them. Moreover, by exploiting variation across multiple conditions, the sensitivity of our approach should only increase with additional data and sources of variation. Hence, the presented framework can likely be applied to test for associations between appropriate continuous quantitative genomic measurements and gene expression, thereby facilitating a comparable basis for meta-analyses on the interplay of epigenetic features.

\section{Materials and Methods}

\subsection{DHS and gene expression data}  The data used in this study were generated through the ENCODE consortium and are publicly available.  Established cell lines and primary cells used in this study were procured from commercial or other sources as listed in Table~\ref{table:Celltypes}. The cells were cultured as per the vendor recommendations, and individual cell growth protocols are available in the UCSC human genome browser. The DHS data are available at the UCSC genome browser by downloading the track IDs listed in Table~\ref{table:DNaseI} and the web address shown therein.  Normalized probe-level expression data were obtained from the Gene Expression Omnibus (GEO); the accession numbers for all arrays are shown in Table~\ref{table:GEO}.  Probes were mapped to genes according to HG18 using bowtie \cite{Langmead:2009:Genome-Biol:19261174} allowing for 2 mismatches and up to 10 maps to the genome, including the best match. Only probe sets for which all probes had a unique best match and fully corresponded to exon boundaries found in RefSeq annotations (HG18) were retained for further analysis. If a RefSeq gene had multiple splice variants, these were aggregated to a meta-gene structure. In the rare event that a gene mapped to currently ambiguous regions (e.g., chr6\_random) such regions were not included. To arrive at a gene specific expression value, the mean expression across all probe sets  within the exon boundaries of the gene model was calculated. This yielded expression measures for 19,215 genes on 20 cell lines. 

\subsection{Statistical methods} The ARS algorithm and statistical analyses were written in the R programming language \cite{R-programming}. The main ARS algorithm, results, GO-analyses, and preprocessed data are available at \\ {\tt http://encode.princeton.edu/}. Complete details of the ARS algorithm, including the null randomization strategy and estimation of the angular penalty, are provided in SI. 

A schematic of the method is shown in Fig.~\ref{fig:method}.  We represented the measurements of a single gene by two paired vectors $\mathbf{x}=(x_1,\ldots,x_m)$ for gene expression and $\mathbf{y}=(y_1,\ldots,y_m)$  for DHS volume, where $m$ is the number of cell-types under consideration (here, $m=20$). To place the two variables on a common scale, each vector was scaled by its maximum observation $\mathbf{x}^s=\frac{\mathbf{x}}{\max\{x_1,\ldots,x_m\}}$ and $\mathbf{y}^s=\frac{\mathbf{y}}{\max\{y_1,\ldots,y_m\}}$ so that all values are now in $[0,1]$.  Each vector was then centered by its median ${\rm med}(\mathbf{x}^s)$ and ${\rm med}(\mathbf{y}^s)$ to form $\mathbf{x}^*=\mathbf{x}^s - {\rm med}(\mathbf{x}^s)$ and $\mathbf{y}^*=\mathbf{y}^s - {\rm med}(\mathbf{y}^s)$. Hence the data for a given gene and segment are now centered around the two-dimensional medoid where the center of mass of the data lies at the origin. If there is little variation across the multiple cell-types, all points would cluster around the medoid, while singular cell-types displaying greater variation would be present as distinct outliers (SI, Fig.~\ref{fig:completeContour} and Fig.~\ref{fig:3Ddensity}). To gauge potential outliers the Euclidian distance $d_i = \sqrt{x_i^{*2}+y_i^{*2}}$ were calculated for every cell type $i=1,\ldots,m$ to produce the distance vector $\mathbf{d}=(d_1,\ldots,d_m)$. We formed a ratio statistic according to $r_i = \frac{d_i}{{\rm med}(\mathbf{d})}$, thereby quantifying the relative distance of each point to the medoid. 

While the ratios $r_i$ describe the dispersion of the data, it does not account for any concordance between the measurements. A perfectly concordant relationship between the two measurements would result in points lying along the $45^\circ$ diagonal identity line. We therefore calculated the angle $\theta_i$ for each data point $(x_i^*, y_i^*)$ relative to the unit vector $(1, 0)$ for $i=1, \ldots, m$, where $0 \leq \theta_i \leq 360$. The angular penalty involves first calculating the smaller of the two angular distances between $\theta_i$ and the identity line, denoted as $\Delta_i$.  For example, $\Delta_i = |45 - \theta_i|$ for $0 \leq \theta_i < 135$.  The angular penalty is calculated as  $a_i = \exp(c \times \Delta_i)$, where $c$ is determined empirically to satisfy a correct null distribution (SI).  Therefore, the value $a_i$ measures the penalized angular distance of $(x_i^*, y_i^*)$ from the identity line in a symmetric fashion (SI, Fig.~\ref{fig:3d-angle}). The statistic applied to each $(x_i^*, y_i^*)$ pair is then ${\rm ARS}_i = a_i \times r_i$, with the gene's overall statistic being the maximum, ${\rm ARS}_{\rm max} = \max({\rm ARS}_1, {\rm ARS}_2, \ldots, {\rm ARS}_m)$.  In addition to calculating these quantities for each gene, we also recorded the ordering of the cell types as determined by their relative ${\rm ARS}_i$ values.

Inclusion of the angular penalty had a twofold purpose. Firstly, it correctly eliminated points that were outliers in only one dimension, gene expression or DHS alone, and therefore not of interest here since there is no direct relationship between the two measurements. Secondly, penalizing such points acted as a tuning parameter adjusting for the degree of off-diagonal noise in data, and thereby ensured a correct null distribution and p-values (Fig.~\ref{fig:pvalueHist}). The specific value of $c$ was determined such that observed null p-values over $(p \geq 0.5)$ had a Uniform(0,1) distribution according to a Kolmogorov-Smirnov test (SI). Reassuringly, this lead to nearly identical values for $c$ across all genomic segment sizes of DHS volume considered (SI, Fig.~\ref{fig:decayScaling}).    

The scaled data $\mathbf{x}^s$ and $\mathbf{y}^s$ for all genes were aggregated into a single distribution in the unit square $[0,1] \times [0,1]$.  From this, randomized data sets were created by sampling 20 points that preserves the fact that either one point must lie on $(1,1)$ or two points lie on $(x,1)$ and $(1,y)$, respectively.  The 20 sampled points are then median centered and the ARS$_{\max}$ statistic is calculated.  We performed this 100 times to generate 100 sets of null ARS$_{\max}$ statistics for every gene (for a total of $100 \times 19,215$ null statistics).  A p-value was then formed for each gene by calculating the frequency by which null statistics exceed the observed statistic. The p-values were then utilized to calculate FDR q-values for the genes, as previously described \cite{Storey:2003:Proc-Natl-Acad-Sci-U-S-A:12883005}. See SI for full details on this randomization method.

\subsection{Selecting local ARS profile peaks for further analysis} We first identified genes called significant at FDR $<$ 0.10 for the ARS analysis performed on the segment size of $\pm$ 200kb about the TSS.  We recorded the maximal cell-type for each of these genes (i.e., the cell type yielding the ARS$_{\max}$ value), producing a list of significant gene/cell-type pairs.  We limited our selection of gene/cell-type pairs to those cell types that were maximal at this threshold for at least 100 genes.  For each of these selected gene/cell-type pairs, we scaled its local ARS profile by the maximal value in the $\pm$ 200kb segment about the TSS.  All DNA sequences $\pm$ 50bp with scaled local ARS profile value $>$ 0.5 were then selected as ``local ARS peaks.''  Likewise, all DNA sequences $\pm$ 50bp with scaled local ARS profile value $<$ 0.2 were selected as the ``negative control set.''  The local ARS peak set consisted of a total of 38,819 100bp regions, and the negative control set consisted of 156,060 100bp regions.  

\subsection{TFBS analysis} We took the above local ARS peaks and negative control set, and we eliminated all segments within $\pm$ 10kb of the TSS,  reducing the number of local ARS peak segments from 38,819 to 32,063 and negative control segments from 156,060 to 148,423. These were searched with all non-redundant vertebrate positions count matrices in the JASPAR database \cite{PortalesCasamar:2010:Nucleic-Acids-Res:19906716}. The position count matrices were converted to position weight matrices using a uniform background, and a matrix specific thresholding of 0.8 of the maximal log-odds score was used. Significant over- or under-representation was determined by exact binomial tests where the probability was based on the frequency of hits per base pair in the negative control sequences. Effect-size was calculated as $\log_2$ fold-change between number of hits per base pair in the local ARS peaks versus the negative control set.

\section*{Web Resource}  To provide an interface for the community to utilize the results from this work, the local ARS tracks across any given gene in any of the 20 cell-types can be calculated via our web-service at {\tt http://encode.princeton.edu/}, where all results encompassing the larger DHS regions are also searchable.

\section*{Acknowledgments}
 We thank the Stamatoyannopoulos lab for useful discussions and suggestions, Shane Neph for help with collating gene ontology analyses, Richard Sandstrom for information on experimental details, Michael Hudock for assistance with computations, and Lance Parsons for building the web site.  The publicly available ENCODE data utilized in this work were generated by the Stamatoyannopoulos lab.  This research was supported in part by NIH grants U54 HG004592 and R01 HG002913.

\end{document}